\documentclass[12pt]{iopart}
\usepackage{graphicx}
\renewcommand{\theequation}{\Roman{section}.\arabic{equation}}
\begin{document}

\title{Relativistic Wave Equations and Compton Scattering}
\author{B.A. Robson and S.H. Sutanto}
\address{Department of Theoretical Physics, Research School of Physical Sciences and Engineering,
The Australian National University, Canberra ACT 0200, Australia}

\begin{abstract}
The recently proposed eight-component relativistic wave equation
is applied to the scattering of a photon from a free electron
(Compton scattering). It is found that in spite of the
considerable difference in the structure of this equation and that
of Dirac  the cross section is given by the Klein-Nishina formula.
\end{abstract}

\section{Introduction}
\label{intro}

Recently an eight-component relativistic wave equation for
spin-$\frac{1}{2}$ particles was
proposed~\cite{RobsonStaudte,Staudte}. This equation was obtained
from a four-component spin-$\frac{1}{2}$ wave equation (the
KG$\frac{1}{2}$ equation~\cite{Staudte}), which contains
second-order derivatives in both space and time, by a procedure
involving a linearization of the time derivative analogous to that
introduced by Feshbach and Villars~\cite{FV} for the Klein-Gordon
equation. This new eight-component equation gives the same
bound-state energy eigenvalue spectra for hydrogenic atoms as the
Dirac equation but has been shown to predict different radiative
transition probabilities for the fine structure of both the Balmer
and Lyman $\alpha$-lines~\cite{Robson}. Since it has been shown
that the new theory does not always give the same results as the
Dirac theory, it is important to consider the validity of the new
equation in the case of other physical problems. One of the early
crucial tests of the Dirac theory was its application to the
scattering of a photon by a free electron : the so-called Compton
scattering problem. In this paper we apply the new theory to the
calculation of Compton scattering to order $e^2$. For this problem
it is easier to calculate the result using the four-component
KG$\frac{1}{2}$ equation and this is carried out in
Section~\ref{4-KG-1/2}. However, for completeness, the calculation
for the eight-component theory is given in Section~\ref{8-theory}
and is shown to be equivalent to that obtained by the
four-component theory, namely the well-known Klein-Nishina
formula~\cite{KleinNishina,Greiner} initially obtained using the
Dirac theory.

\section{Four-component (KG$\frac{1}{2}$ equation) theory for Compton scattering}
\label{4-KG-1/2}

The four-component KG$\frac{1}{2}$ equation~\cite{Staudte} is
\begin{equation}
\left[ \left( D_\mu D^{\mu} + m^2 \right) \mbox{\boldmath $1_4$} +
\frac{e}{2} \sigma^{\mu\nu} F_{\mu \nu}\right] \Psi = 0
\label{4component}
\end{equation}
where $\sigma^{\mu \nu} = \frac{i}{2}
\left[\gamma^{\mu},\gamma^{\nu}\right]$, $F_{\mu \nu} =
\partial_\mu A_\nu - \partial_\nu A_\mu$ and $\gamma^\mu$ are the
standard Dirac matrices :

\begin{equation}
\begin{array}{lr} \gamma^0 = \left[\begin{array}{cc} \mbox{\boldmath $1_2$} & \mbox{\boldmath $0$} \\ \mbox{\boldmath $0$} & \mbox{\boldmath $-1_2$} \end{array} \right]      &        \mbox{\boldmath $\gamma$} = \left[\begin{array}{cc} \mbox{\boldmath $0$} & \mbox{\boldmath $\sigma$} \\ \mbox{\boldmath $-\sigma$} & \mbox{\boldmath $0$} \end{array} \right] \end{array}.     \label{gamma}
\end{equation}
The final term is the spin interaction term in the presence of an
external electromagnetic field. Eq.~(\ref{4component}) reduces to
the free particle Klein-Gordon type equation in the absence of a
field
\begin{equation}
\left\{ \partial_\mu \partial^\mu + m^2 \right\} \mbox{\boldmath
$1_4$} \Psi = 0 .   \label{KleinGordon}
\end{equation}

\begin{figure}
\includegraphics[scale=0.8]{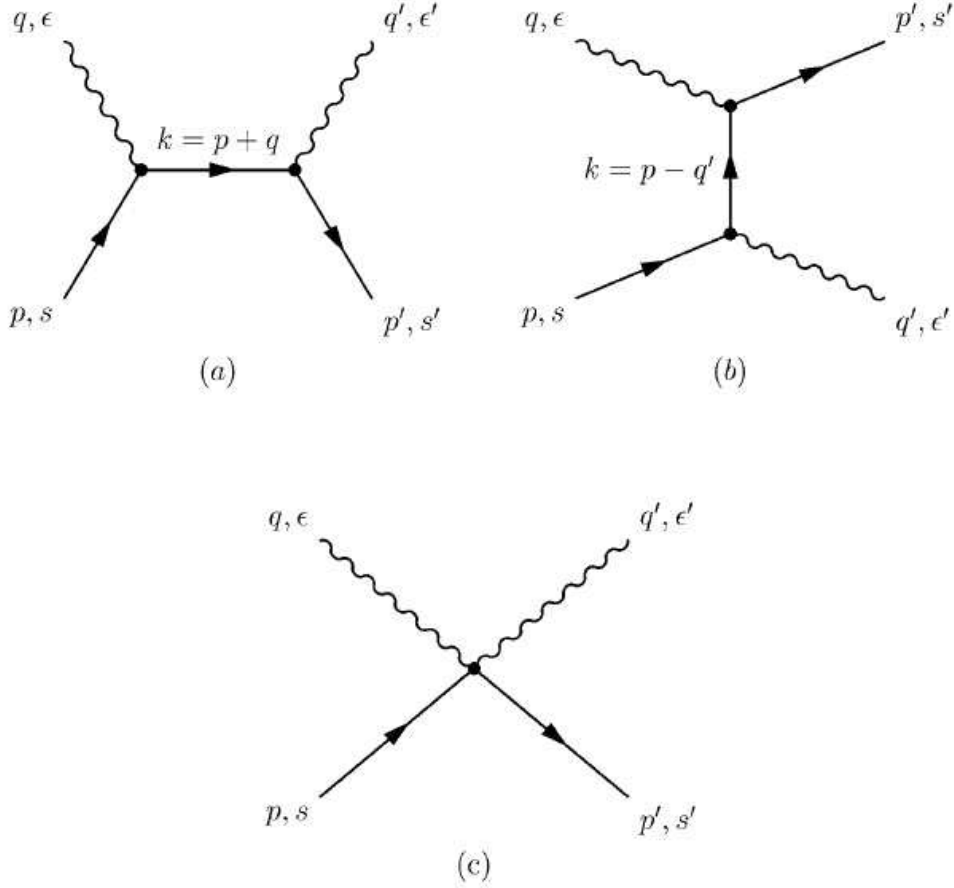}
\caption{Feynman diagrams for Compton scattering.} \label{fig-1}
\end{figure}

The free particle positive energy solution of (\ref{KleinGordon}),
normalized within a box of volume $L^3$ may be written
\begin{equation}
\Psi = \frac{e^{-ip \cdot x}}{\sqrt{2EL^3}} u(p,s)
\label{solusi}
\end{equation}
where
\begin{equation}
u(p,s) = \frac{1}{\sqrt{2mE_+}} \left[ \begin{array}{c} E_+ \phi_s
\\ \left( \mbox{\boldmath $\sigma \cdot p$} \right) \phi_s
\end{array} \right] .  \label{spinor}
\end{equation}
Here $p$ is the four-momentum of the electron of energy $E$, $E_+$
= $E$ + $m$ and $\phi_s$ is a normalized two component spinor in
the rest frame.
Compton scattering refers to the scattering of photons by free
electrons. The lowest order Feynman diagrams for this process are
shown in Figure~\ref{fig-1}. There are three diagrams : (a)
direct, (b) exchange and (c) ``seagull''.

The perturbing interaction causing the scattering is given by
\begin{equation}
V(x) = ie\left(\partial_\mu A^\mu +A_\mu \partial^\mu \right) +
\frac{e}{2} \sigma^{\mu \nu} F_{\mu \nu} - e^2 A^\mu A_\mu
\label{potential}
\end{equation}
where $A_\mu$ is the four-vector electromagnetic potential
satisfying
\begin{equation}
\partial_\mu \partial^\mu A_\nu = 0. \label{Lorentzcondition}
\end{equation}
The direct and exchange diagrams correspond to the first and
second terms in the potential $V(x)$. The ``seagull'' diagram
arises from the third term in Eq.~(\ref{potential}).

To describe an incoming (outgoing) photon of polarization vector
$\epsilon_\mu$ ($\epsilon_\mu^{\prime}$) we choose the plane wave
solutions of (\ref{Lorentzcondition}) to be of the form

\begin{equation}
A^\mu (x,q) = \frac{1}{\sqrt{2q^0 L^3}} \epsilon^\mu (q,\lambda)
e^{-iq\cdot x} \label{4vecpot}
\end{equation}
and
\begin{equation}
A^{\mu} (x,q^\prime) = \frac{1}{\sqrt{2q^{\prime 0} L^3}}
\epsilon^{\prime \mu}(q^\prime,\lambda^\prime)
e^{+iq^{\prime}\cdot x} \label{4vecpotprime}
\end{equation}
respectively, where $q$, $q^\prime$ are the four-momenta of the
photons,  $q^2$ = $q^{\prime 2}$ = 0 and $\lambda ,
\lambda^\prime$ define the photon polarization states.

The differential cross section for Compton scattering is given by
\begin{equation}
d \sigma = \int \frac{\left| a_{fi} \right|^2}{\left|
\mbox{\boldmath$j$}\right| T }dN \label{diffcrosssection}
\end{equation}
where $\left| a_{fi} \right| ^2$ is the transition probability,
$T$ is the time interval,  $\mbox{\boldmath$j$}$ is the incident
particle flux, and $N$ is the number of final states. The
transition matrix from standard pertubation theory may be written
\begin{eqnarray}
a_{fi} & = & i \int d^4 x  \bar{\Psi}_f (x) V(x) \Psi_i (x)
\nonumber \\
          & + & i \int \int d^4 x d^4 y  \bar{\Psi}_f (x) V(x) G(x-y) V(y) \Psi_i (y) \nonumber \\
          & + & ... \label{transitionmatrix}
\end{eqnarray}
where $\bar{\Psi}$ = $\Psi^\dagger \gamma_0$ and the Green
function
\begin{equation}
G(x-y) = \int{\frac{d^4 k}{(2\pi)^4} e^{-ik\cdot(x-y)} G(k)}
\label{greenfunction}
\end{equation}
with

\begin{equation}
G(k) = \frac{1}{k^2 - m^2} \mbox{\boldmath$1_4$}.
\label{propagator}
\end{equation}

To the lowest order in e, the transition matrix for Compton
scattering is, from Eqs.~(\ref{potential}) and
(\ref{transitionmatrix})

\begin{eqnarray}
a_{fi} & = & - i e^2 \int d^4x \bar{\Psi}_f (x) [A^\mu
(x,q^\prime) A_\mu (x,q) + A^\mu (x,q) A_\mu (x,q^\prime)] \Psi_i
(x) \nonumber \\
       & - & i e^2 \int \int d^4x d^4y \bar{\Psi}_f \left[ \left( \partial^\mu A_\mu + A_\mu \partial^\mu - \frac{i}{2} \sigma^{\mu \nu} F_{\mu \nu} \right)_{(x,q)} G(x - y) \right. \nonumber \\
       & \times & \left. \left( \partial^\mu A_\mu + A_\mu \partial^\mu -\frac{i}{2} \sigma^{\mu \nu} F_{\mu \nu} \right)_{(y,q^\prime)} \right. \nonumber \\
       & + & \left. \left( \partial^\mu A_\mu + A_\mu \partial^\mu -\frac{i}{2} \sigma^{\mu \nu} F_{\mu \nu} \right)_{(x,q^\prime)} G (x - y) \right. \nonumber \\
       & \times & \left. \left( \partial^\mu A_\mu + A_\mu \partial^\mu -\frac{i}{2} \sigma^{\mu \nu} F_{\mu \nu} \right)_{(y,q)} \right] \Psi_i (y)   \label{transmat1}
\end{eqnarray}
where the bracketed subscripts $(x,q)$ denote the arguments of
$\partial_\mu$ and  $A_\mu$.

To evaluate Eq.~(\ref{transmat1}) it is convenient to choose the
laboratory frame with the target electron at rest, for which

\begin{eqnarray}
p_\mu = (m, {\bf 0})\hspace{7mm}q_\mu = (\omega,{\bf
q})\hspace{5mm}\mbox{with}\hspace{5mm}\omega = \left|{\bf
q}\right| \nonumber \\
p^\prime_\mu = (E^\prime,{\bf
p}^{\prime})\hspace{7mm}q_\mu^{\prime} = (\omega^\prime, {\bf
q}^\prime)\hspace{3mm}\mbox{with}\hspace{4mm}\omega^{\prime} =
\left|\bf{q}^\prime\right|   \label{momenta}
\end{eqnarray}
and the special gauge in which the initial and final photons are
transversely polarized in laboratory frame :

\begin{eqnarray}
\epsilon_\mu = (0,
\mbox{\boldmath$\epsilon$})\hspace{5mm}\mbox{with}\hspace{5mm}\mbox{\boldmath$\epsilon
\cdot q$} = 0 \nonumber \\
\epsilon_{\mu}^\prime = (0,
\mbox{\boldmath$\epsilon^\prime$})\hspace{4mm}\mbox{with}\hspace{5mm}\mbox{\boldmath$\epsilon^\prime
\cdot q^\prime$} = 0. \label{CoulombGauge}
\end{eqnarray}

Inserting Eqs.~(\ref{solusi}), (\ref{4vecpot}),
(\ref{4vecpotprime}), and (\ref{greenfunction}) into
Eq.~(\ref{transmat1}) and carrying out the integration over $d^4
y$, $d^4 k$ and $d^4 x$ gives

\begin{equation}
a_{fi}  =  \frac{i e^2 (2\pi)^4}{\sqrt{2E2E^\prime 2\omega
2\omega^\prime L^{12}}} \delta^4 (p^\prime + q^\prime - p - q)
\left[ \bar{u} (p^\prime , s^\prime) \Gamma  u(p,s) \right]
\label{transmat2}
\end{equation}
where

\begin{equation}
\Gamma = \frac{(2p^\prime \cdot \epsilon^\prime - \not\!q^\prime
\not\! \epsilon^\prime) \not\! q \not\! \epsilon}{2p \cdot q} +
\frac{(2p^\prime \cdot \epsilon + \not\! q \not\! \epsilon) \not\!
q^\prime \not\! \epsilon^\prime}{2p \cdot q^\prime} - 2 (\epsilon
\cdot \epsilon^\prime) \mbox{\boldmath$1_4$}.
\label{Gammafunction}
\end{equation}
It should be noted that the factor of two in the last term
involving $\epsilon^\prime \cdot \epsilon$ arises from two equal
contributions to the seagull diagram (i.e. $\epsilon^\prime \cdot
\epsilon$ +  $\epsilon \cdot \epsilon^\prime$). In
(\ref{Gammafunction}) the slash notation $\not\!\! A$ = $\gamma^0
A^0$ - \mbox{\boldmath$\gamma \cdot A$} has been used.

Using Eqs.~(\ref{diffcrosssection}) and (\ref{CoulombGauge}) and
the Compton relation

\begin{equation}
\frac{\omega}{\omega^\prime} = 1 + \frac{\omega}{m}(1 - \cos
\theta) \label{compton}
\end{equation}
where the final photon is scattered by an angle $\theta$ into the
spherical angle element $d \Omega_{q^\prime}$ with respect to the
incident photon, the differential cross section is

\begin{equation}
\frac{d \sigma}{d \Omega_{q^\prime}}(\lambda^\prime, \lambda;
s^\prime, s) = \alpha^2 \frac{\omega^{\prime 2}}{\omega^2}
\frac{1}{4m^2}  \left| \bar{u} (p^\prime, s^\prime) \Gamma  u(p,s)
\right|^2 .  \label{transmat3}
\end{equation}
Here $\alpha$ is the fine structure constant $e^2/4 \pi$ and
$\lambda, \lambda^\prime$ are the initial and final photon
polarizations.

Averaging over the initial electron spins and summing over the
final electron spins gives

\begin{equation}
\frac{d \bar{\sigma}}{d \Omega_{q^\prime}}(\lambda^\prime,
\lambda) = \frac{1}{2}\sum_{s^\prime,s} \frac{d \sigma}{d
\Omega_{q^\prime}}(\lambda^\prime, \lambda; s^\prime, s).
\label{spinsum}
\end{equation}
The electron spinors can be eliminated employing the usual trace
techniques~\cite{Aitchison} and Eq.~(\ref{spinsum}) gives

\begin{equation}
\frac{d \bar{\sigma}}{d \Omega_{q^\prime}}(\lambda^\prime,
\lambda) = \alpha^2 \frac{\omega^{\prime 2}}{\omega^2}
\frac{1}{8m^2}Tr \left[ \frac{\not\! p^\prime + m}{2m} \Gamma
\frac{\not\! p + m}{2m}\bar{\Gamma} \right]      \label{transmat4}
\end{equation}
where

\begin{equation}
\bar{\Gamma} = \gamma_0 \Gamma^{\dagger} \gamma_0.
\end{equation}

The calculation of the trace in Eq.~(\ref{transmat4}) is rather
tedious, involving products of up to ten $\gamma$ matrices.
However, the result is quite simple :

\begin{equation}
Tr \left[ \frac{\not\! p^\prime + m}{2m} \Gamma \frac{\not\! p +
m}{2m}\bar{\Gamma} \right] = 8 ( \epsilon \cdot \epsilon^\prime)^2
+ \frac{2(q \cdot q^\prime)^2}{(p \cdot q)(p \cdot q^\prime)}.
\label{trace}
\end{equation}
Now $(q \cdot q^\prime) = \omega \omega^\prime (1 - \cos \theta)$,
$(p \cdot q) = m \omega$ and $(p \cdot q^\prime) = m
\omega^\prime$ so that

\begin{equation}
\frac{d \bar{\sigma}}{d \Omega_{q^\prime}}(\lambda,
\lambda^\prime) = \frac{\alpha^2}{2m^2} \left(
\frac{\omega^\prime}{\omega} \right)  \left[4 (\epsilon \cdot
\epsilon^\prime)^2 + \frac{\omega \omega^\prime}{m^2}(1 - \cos
\theta)^2 \right]      \label{KleinNishina}
\end{equation}
which is identical to the Klein-Nishina formula derived using the
Dirac equation~\cite{KleinNishina,Greiner}.

\section{Eight-component theory for Compton scattering}
\label{8-theory}

The eight-component relativistic wave equation is
obtained~\cite{RobsonStaudte} from the KG$\frac{1}{2}$ equation by
linearizing the time derivative using a procedure analogous to
that employed by Feshbach and Villars~\cite{FV} for the
Klein-Gordon equation. The eight-component (FV$\frac{1}{2}$
equation) has the following form

\begin{equation}
i \frac{\partial}{\partial t}\mbox{\boldmath$1_8$} \Psi_{FV
\frac{1}{2}} = H_{FV \frac{1}{2}} \Psi_{FV \frac{1}{2}}
\label{8component1}
\end{equation}
where
\begin{equation}
H_{FV \frac{1}{2}}  =  (\tau_3 + i \tau_2) \otimes
\frac{1}{2m}\left[- \mbox{\boldmath$D$}^2 \mbox{\boldmath$1_4$} +
\frac{e}{2} \sigma^{\mu \nu} F_{\mu \nu} \right] + m (\tau_3
\otimes \mbox{\boldmath$1_4$}) + e A_0 \mbox{\boldmath$1_8$}.
\label{hamiltonian}
\end{equation}
In Eq.~(\ref{hamiltonian}) $\tau_i$ are the standard Pauli
matrices and $\otimes$ is the usual Kronecker (direct) product.

The free positive energy electron solution of (\ref{8component1}),
normalized within a box of volume $L^3$, may be written
\begin{equation}
\Psi_{FV \frac{1}{2}} = \frac{1}{2m} \left( \begin{array}{c} E_+
\\ -E_- \end{array}\right) e^{-i p \cdot x} \otimes u(p,s)
\label{normalizedsolution}
\end{equation}
where $E_\pm = E \pm m$ and the inner product can be
written~\cite{Staudte} as

\begin{equation}
\left<\Psi_{FV \frac{1}{2}}| \Psi_{FV \frac{1}{2}}\right> = \int
\Psi_{FV \frac{1}{2}}^{\dagger} (x) \tau_4 \Psi_{FV \frac{1}{2}}
(x) d^3x            \label{innerproduct}
\end{equation}
where $\tau_4 = \tau_3 \otimes \gamma_0$.

The differential cross section for Compton scattering is given by
Eq.~(\ref{diffcrosssection}). In the eight-component theory, the
transition matrix is

\begin{eqnarray}
a_{fi} & = & i \int d^4x \Psi_f^{\dagger} (x) \tau_4 V (x) \Psi_i
(x) \nonumber \\
      & + & i \int \int d^4x d^4y \Psi_f^{\dagger} (x) \tau_4 V (x) G (x - y) V (y) \Psi_i (x) \nonumber \\
      & + & ...                        \label{transitionmatrix2}
\end{eqnarray}
where now the Green function $G (x - y)$ is given by
Eq.~(\ref{greenfunction}) but with
\begin{equation}
G (k) = \frac{1}{k^2 - m^2} \left[ \begin{array}{cc}
\frac{\mbox{\boldmath$k$}^2}{2m} + m + k_0 &
\frac{\mbox{\boldmath$k$}^2}{2m} \\
-\frac{\mbox{\boldmath$k$}^2}{2m} &
-\frac{\mbox{\boldmath$k$}^2}{2m} - m + k_0 \end{array} \right]
\otimes \mbox{\boldmath$1_4$}     \label{propagator2}
\end{equation}
The perturbing interaction is

\begin{eqnarray}
V (x) & = & \frac{1}{2m}(\tau_3 + i \tau_2) \nonumber \\
         & \otimes &\left[ie \left(\{\mbox{\boldmath$\nabla \cdot A$} + \mbox{\boldmath$A \cdot \nabla$}\} \bf{1_4} \right) + \frac{e}{2} \sigma^{\mu \nu} F_{\mu \nu} + e^2 (\mbox{\boldmath$A \cdot A$}) \bf{1_4}  \right].        \label{potential2}
\end{eqnarray}

Inserting (\ref{potential2}) into Eq.~(\ref{transitionmatrix2})
gives to second order in $e$

\begin{eqnarray}
a_{fi} & = & \frac{i e^2}{2m} \int d^4x \Psi_f^{\dagger} (x)
\tau_4 (\tau_3 + i \tau_2) \otimes \mbox{\boldmath$A$} (x,q) \cdot
\mbox{\boldmath$A$} (x,q^\prime) \Psi_i (x) \nonumber \\
      & - & \frac{i e^2}{4m^2} \int \int d^4x d^4y \Psi_f^{\dagger} (x) \nonumber \\
      & \times & \tau_4 \left[ (\tau_3 +i \tau_2) \otimes \left((\not\! \partial \not\!\! A) + \left\{\mbox{\boldmath$\nabla \cdot A$} + \mbox{\boldmath$\nabla \cdot A$} \right\} \bf{1_4}\right) \right]_{(x,q)} G (x - y) \nonumber \\
      & \times & \tau_4 \left[ (\tau_3 +i \tau_2) \otimes \left((\not\! \partial \not\!\! A) + \left\{\mbox{\boldmath$\nabla \cdot A$} + \mbox{\boldmath$\nabla \cdot A$} \right\} \bf{1_4}\right) \right]_{(y,q^\prime)} \nonumber \\
      & + & \tau_4 \left[ (\tau_3 +i \tau_2) \otimes \left((\not\! \partial \not\!\! A) + \left\{\mbox{\boldmath$\nabla \cdot A$} + \mbox{\boldmath$\nabla \cdot A$} \right\} \bf{1_4}\right) \right]_{(x,q^\prime)} G (x - y) \nonumber \\
      & \times & \tau_4 \left[ (\tau_3 +i \tau_2) \otimes \left((\not\! \partial \not\!\! A) + \left\{\mbox{\boldmath$\nabla \cdot A$} + \mbox{\boldmath$\nabla \cdot A$} \right\} \bf{1_4}\right) \right]_{(y,q)} \Psi_i (y)
  \label{transmat5}
\end{eqnarray}
where the use of brackets in \mbox{$\not\! \partial \not\!\! A$}
will mean that \mbox{$\not\!\partial$} operates only within the
brackets.

As already seen in Section~\ref{4-KG-1/2}, to second order in $e$,
the amplitude for Compton scattering involves the three Feynman
diagrams shown in Figure~\ref{fig-1}. Evaluation of
Eq.~(\ref{transmat5}) choosing the laboratory frame with the target
electron at rest and the special transverse gauge in which the
initial and final photons are transversely polarized in laboratory
frames gives

\begin{eqnarray}
a_{fi}& = & \frac{i e^2 (2 \pi)^4}{8m^3 \sqrt{2 \omega 2
\omega^\prime L^{12}}} \delta^4 (p^\prime + q^\prime - p - q)
(E_+, -E_-) \nonumber \\
         & \otimes &  u^\dagger (p^\prime, s^\prime) \tau_4 [ \frac{1}{2 p \cdot q} ((\tau_3 + i\tau_2) \otimes (-2 \mbox{\boldmath$p^\prime \cdot \epsilon^\prime$} - \not\! q^\prime \not\! \epsilon^\prime) \not\! q \not\! \epsilon)   \nonumber \\
         & + &  \frac{1}{2p \cdot q^\prime}((\tau_3 + i \tau_2) \otimes (-2 \mbox{\boldmath$p^\prime \cdot \epsilon$} + \not\! q \not\! \epsilon) \not\! q^\prime \not\! \epsilon^\prime)  \nonumber \\
         & + & 2 (\mbox{\boldmath$\epsilon \cdot \epsilon^\prime$})(\tau_3 + i \tau_2) \otimes \mbox{\boldmath$1_4$}] \left( \begin{array}{c} E_+ \\ -E_- \end{array} \right) \otimes u(p,s).     \label{transmat6}
\end{eqnarray}
Using 4 x 4 block matrices, it is readily shown that the
expression (\ref{transmat6}) reduces to

\begin{equation}
a_{fi} = \frac{i e^2 (2 \pi)^4}{2m \sqrt{2 \omega 2 \omega^\prime
L^{12}}} \delta^4 (p^\prime + q^\prime - p - q) \left\{ u^\dagger
(p^\prime, s^\prime) \gamma^0 \Gamma u(p,s) \right\}
\label{transmat7}
\end{equation}
where
\begin{eqnarray}
\Gamma & = & \left[ \frac{1}{2 p \cdot q} ((-2
\mbox{\boldmath$p^\prime \cdot \epsilon^\prime$} - \not\! q^\prime
\not\! \epsilon^\prime) \not\! q \not\! \epsilon)  \right.
\nonumber \\
         & + & \left.  \frac{1}{2p \cdot q^\prime}((-2 \mbox{\boldmath$p^\prime \cdot \epsilon$} + \not\! q \not\! \epsilon) \not\! q^\prime \not\! \epsilon^\prime) + 2 (\mbox{\boldmath$\epsilon \cdot \epsilon^\prime$}) \bf{1_4} \right].  \label{Gamma2}
\end{eqnarray}
For the special transverse gauge (\ref{CoulombGauge}),
Eq.~(\ref{Gamma2}) is identical with Eq.~(\ref{Gammafunction}) so
that the differential cross section for Compton scattering is once
again the Klein-Nishina formula of Eq.~(\ref{KleinNishina}).

\section{Conclusion}
\label{conclusion}

It has been shown that to order $e^2$ both the second order
KG$\frac{1}{2}$ equation and its eight-component form give the
Klein-Nishina formula for the Compton scattering problem. This is
the same result as obtained by the standard Dirac theory, although
the new theory involves an additional ``seagull" Feynman diagram.


\end{document}